\documentclass[useAMS,usenatbib]{mn2e}
\usepackage{txfonts,graphicx,natbib}
\bibpunct{(}{)}{;}{a}{}{,}
\begin{document}

\title[The transitional SN 2005la]{Massive stars exploding in a He-rich circumstellar medium.  \\
    II. The transitional case of SN 2005la}

\author[Pastorello et al.]{A. Pastorello$^{1}$\thanks{E-mail:a.pastorello@qub.ac.uk}, 
R. M. Quimby$^{2,3}$, S. J. Smartt$^{1}$, S. Mattila$^{1,4}$, H. Navasardyan$^{5}$, \and R. M. Crockett$^{1}$, N.
Elias-Rosa$^{5,6}$, P. Mondol$^{2}$, J. C. Wheeler$^{2}$, 
D. Young$^{1}$\\
$^{1}$ Astrophysics Research Centre, School of Mathematics and Physics, 
Queen's University Belfast, Belfast BT7 1NN, United Kingdom\\
$^{2}$ Department of Astronomy, University of Texas, Austin, TX 78712, US\\
$^{3}$ Current address: California Institute of Technology, Pasadena, CA 91125, US\\
$^{4}$ Tuorla Observatory, University of Turku, V\"ais\"al\"antie 20, FI-21500 Piikki\"o, Finland \\
$^{5}$ INAF Osservatorio Astronomico di Padova, Vicolo dell'Osservatorio 5, 35122 Padova, Italy \\
$^{6}$ Max-Planck-Institut f\"ur Astrophysik, Karl-Schwarzschild-Str. 1, D-85741 Garching bei M\"unchen, Germany}

\date{Accepted .....; Received .....; in original form .....}


\maketitle

\label{firstpage}

\begin{abstract}
We present photometric and spectroscopic data of the peculiar SN~2005la, an object
which shows an optical light curve with some luminosity fluctuations and spectra
with comparably strong narrow hydrogen and helium lines, probably of circumstellar nature. 
The increasing full-width-half-maximum velocity of these lines is indicative of an 
acceleration of the circumstellar material. SN~2005la exhibits hybrid
properties, sharing some similarities with both type IIn supernovae and 2006jc-like (type Ibn)
events. We propose that the progenitor of SN~2005la was a very young Wolf-Rayet (WN-type) star 
which experimented mass ejection episodes shortly before core collapse. 
\end{abstract}


\section{Introduction} \label{intro}

The interest of the astronomical community has been recently 
caught by a few very luminous supernovae (SNe) that have exploded within dense, massive and
pre-existent circumstellar media. Some of them have been spectroscopically classified as type 
IIn supernovae 
\citep[SNe IIn,][]{sch90} because of the presence of prominent narrow (about 1000 km s$^{-1}$) H Balmer emission lines 
(e.g. SN 2006gy, \citet{smi06,ofek07}; SN 2007bw, \citet{NSF07}). For few 
other events (e.g. SN 1999cq, \citet{mat00}; SN 2006jc, \citet{fol07,pasto07,pasto07b}), whose
spectra show relatively narrow He I emission lines ($\sim$ 2000 km s$^{-1}$), a different nomenclature was coined: type Ibn SNe
\citep{pasto07,pasto07b}.
In the spectra of these objects, the narrow H lines are much weaker than the He I features, or completely missing.

In addition, a number of regular type IIn SNe have developed at some stages clear evidence of narrow (from a few $\times$100 km s$^{-1}$
to $\sim$2000 km s$^{-1}$)
He I lines although, 
contrary to SNe Ibn, they are always less prominent than H$\alpha$. This group includes SN 1978K \citep{chu95}, 
SN 1987B \citep{sch96}, SN 1988Z \citep{sta91,tur93,are99},
SN 1995G \citep{pasto02}, SN 1995N \citep{fran02,pasto05,zam05}, SN 1996L \citep{ben99}, SN 1997ab \citep{heg97},
SN 1997eg \citep{sal02}, SN 1998S \citep{leo00,ger00,liu00,fas01,anu01}.
In all these SN events, the narrow lines (both of H and He I) are thought to originate in a slowly-moving 
circumstellar medium (CSM) which is photoionized either by early UV flux or interaction with the SN ejecta. 

A number of authors \citep{kot06,gal07,smi06} suggest that a significant fraction of SNe IIn might be 
produced by the explosion of massive Luminous Blue Variable
(LBV) stars, even if the nature of at least some of type IIn events is still debated 
\citep[e.g. SN~2002ic and similar events, see][and references therein]{ham03,ben06,pri07a,tru08}.
As an alternative view, \cite{woo07} propose a pair-instability pulsation scenario as a trigger of 
major mass ejection episodes. The ejecta collide with the material expelled by the star 
in previous eruptions converting large amounts of kinetic energy into radiation and generating a Type IIn SN.

Type Ibn SNe seem to be connected with WR progenitors whose He envelope is almost completely stripped away, and that 
explode while they are still embedded within the He-rich material \citep{pasto07,pasto07b,fol07,smi07,tom07,imm07}.
\citet{pasto07} showed that SN 2006jc was preceded by a giant outburst 2 years before core-collapse. 
Although the SN itself suggests a WC star progenitor, the outburst had similar characteristics to an
LBV eruption.

In \citet[][paper 1 of this series]{pasto07b} the common properties 
of objects similar to SN~2006jc (SNe Ibn) are discussed. In paper 3 \citep{seppo07} we investigate
the origin of the strong infrared excess observed in SN 2006jc.
In this paper (paper 2) we describe the observed characteristics
of the unusual SN~2005la, an object whose spectra show simultaneously narrow H and He I features, 
with the most prominent He I lines having comparable strength (at least at early stages) as H$\alpha$, 
and showing unusual evolution with time. 
We propose that SN~2005la is a transitional object lying somewhere between the newly defined SNe Ibn 
and more classical type IIn events.

\section{An Unprecedented Supernova} \label{general}

\begin{figure}
\centering
\includegraphics[width=8.5cm]{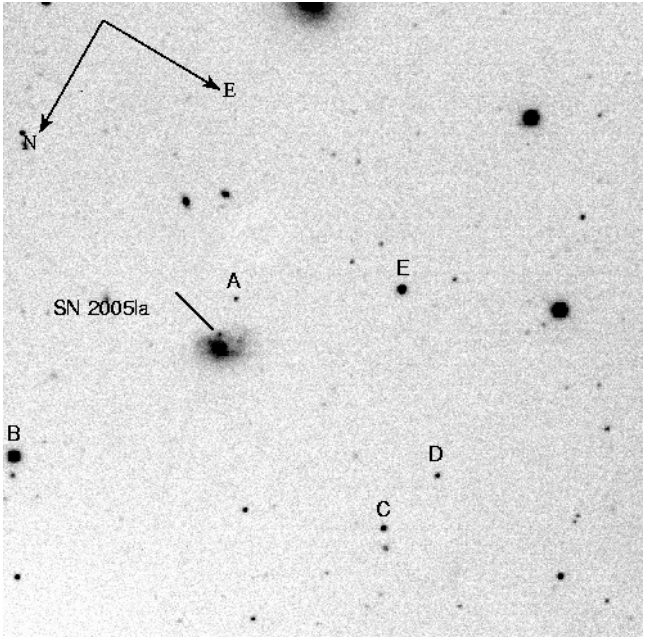}
\caption{The field of SN~2005la with the local sequence of stars labelled
with capital letters. \label{fig1}}
\end{figure}

\begin{table}
\centering
\caption{Sequence of local standards in the field of KUG 1249+278. The errors reported in brackets are
the r.m.s. of the derived average magnitudes.} \label{tab1}
\begin{tabular}{ccccc} \\
\hline
Star & B band & V band & R band & I band \\
\hline \hline
A & 19.96 (0.01) & 19.64 (0.01) & 19.34 (0.01) & 18.32 (0.01) \\
B & 15.63 (0.01) & 14.79 (0.01) & 14.22 (0.01) & 13.86 (0.01) \\
C & 19.92 (0.02) & 18.61 (0.01) & 17.87 (0.01) & 17.16 (0.02) \\
D & 21.46 (0.04) & 19.93 (0.02) & 18.83 (0.01) & 17.57 (0.01) \\ 
E & 16.94 (0.01) & 16.26 (0.02) & 16.00 (0.01) & 15.58 (0.02) \\
\hline
\end{tabular}
\end{table}

SN 2005la  was discovered in the spiral galaxy KUG 1249+278
by R. Q. and P. M. on behalf of the Texas Supernova 
Search\footnote{http://grad40.as.utexas.edu/$\sim$quimby/tss/index.html}. 
The unfiltered discovery images were obtained
on Nov. 30.51 and Dec. 1.49 UT with the 0.45-m ROTSE-IIIb telescope 
at the McDonald Observatory, in which the SN appeared to be at a magnitude 
of about 17.6 \citep{puc05}. 
The coordinates of the new object were
R.A. = $12^{h}52^{m}15\fs68$, Decl. = $+27\degr31\arcmin52\farcs5$ (equinox 2000.0; 
uncertainty +/- 0$\farcs$8), 6$\arcsec$ W and 6$\arcsec$ S of the center of the host galaxy. 
The field of KUG 1249+278 with SN~2005la is shown in Fig. \ref{fig1},
where 5 reference stars are also labelled.

SN~2005la is hosted in an Sc-type galaxy, with total apparent B band magnitude
16.67$\pm$0.50 (LEDA ~database)\footnotemark.
\footnotetext{http://leda.univ-lyon1.fr/}
The distance to SN~2005la can be determined from the recession velocity
of the host galaxy, which is derived from the position of the narrow H$\alpha$ emission 
line of a nearby H II region \citep[v$_{rec}$ = 5570 km s$^{-1}$,][]{fil05}. It is  
estimated to be about 78.5 Mpc (obtained adopting H$_0$ = 71 km s$^{-1}$ Mpc$^{-1}$), 
providing a distance modulus of $\mu$ = 34.47.
Assuming a negligible Galactic extinction (see below), we obtain for KUG 1249+278 a 
total absolute magnitude M$_B \approx$  -17.8.
The host galaxy of SN~2005la is therefore a relatively faint spiral, similar to
the galaxies hosting SNe~2006jc and 2002ao \citep[LEDA, see also][]{pri07b}.
An average metallicity estimate for KUG 1249+278 can be obtained via the 
best fit of the luminosity-metallicity relation of \citet{tre04}. 
This provides 
12+log(O/H) = 8.53 $\pm$ 0.16 (dex) for the above estimated galaxy absolute 
magnitude. We can adopt this value as only an upper limit for the oxygen abundance at 
the SN location, because we presume there are some abundance gradients throughout
 KUG 1249+278 (although probably rather modest due to the characteristics of the host galaxy). 
Interestingly, the derived abundance is very
similar to those estimated for three (out of four)\footnotemark
\footnotetext{A similar low-metallicity value was not found in the case of the host galaxy of the 
type Ibn SN 1999cq, for which a significantly higher oxygen abundance was computed, i.e.
12+log(O/H) $\approx$ 9.3 $\pm$ 0.2 \protect\citep{pasto07b}. However, abundance gradients throughout 
the galaxy were not taken into account in that estimate, and in such a luminous spiral galaxy
they are expected to be relatively large.}
 SNe Ibn analysed in \citet{pasto07b}.

\citet{fil05}, on the basis of a spectrum obtained on Dec. 3 UT with the
Keck II 10-m telescope (+ DEIMOS), classified this SN as a peculiar type
II/Ib event, having weak and narrow H and He I lines with P-Cygni profiles,
but dominated by the component in emission.
The line velocity derived both from the position of the absorption component
minimum and the full-width-half-maximum (FWHM) of the emission component 
was about 1700-1800 km s$^{-1}$.

The low signal-to-noise spectra of SN~2005la (see Sect. \ref{spec}) do not
allow to provide reliable estimates for the extinction within the host galaxy.
However SN~2005la has quite a peripheral location and 
the extinction suffered by the SN is expected to be rather small.
\citet{sch98} provide a very small value for the Galactic extinction in the direction
of the SN, being E(B-V) = 0.011 magnitudes. This value will be adopted throughout 
the paper as an estimate of the total extinction.

\section{Light and colour Curves} \label{photo}

\begin{table*}
\centering
\small
\caption{Photometry of SN~2005la, including the most significant detection limits. \label{tab2}} 
\begin{tabular}{cccccccc} \\
\hline \hline
Date & JD & Days from maximum & B band & V band & R band & I band & Instrument \\
\hline
Nov 19, 2005 & 2453693.62& -0.9 & & & $>$17.50 & & ROTSE-IIId \\
Nov 19, 2005 & 2453693.99& -0.5 & & & $>$16.75 & & ROTSE-IIIb \\
Nov 20, 2005 & 2453694.99& +0.5 & & & $>$17.17 & & ROTSE-IIIb \\
Nov 21, 2005 & 2453695.99& +1.5 & & & 17.33 (0.08) & & ROTSE-IIIb \\
Nov 22, 2005 & 2453696.99& +2.5 & & & 17.39 (0.10) & & ROTSE-IIIb \\
Nov 23, 2005 & 2453697.99& +3.5 & & & 17.42 (0.09) & & ROTSE-IIIb \\
Nov 29, 2005 & 2453704.00& +9.5 & & & 17.68 (0.08) & & ROTSE-IIIb \\
Nov 30, 2005 & 2453705.01& +10.5 & & & 17.59 (0.09) & & ROTSE-IIIb \\
Dec 01, 2005 & 2453705.60& +11.1 & & & 17.52 (0.13) & & ROTSE-IIId \\
Dec 01, 2005 & 2453705.99& +11.5 & & & 17.53 (0.08) & & ROTSE-IIIb \\
Dec 02, 2005 & 2453706.99& +12.5 & & & 17.55 (0.07) & & ROTSE-IIIb \\
Dec 03, 2005 & 2453707.99& +13.5 & & & 17.64 (0.08) & & ROTSE-IIIb \\
Dec 04, 2005 & 2453708.99& +14.5 & & & 17.72 (0.09) & & ROTSE-IIIb \\
Dec 05, 2005 & 2453709.59& +15.1 & & & 17.73 (0.11) & & ROTSE-IIId \\
Dec 05, 2005 & 2453709.99& +15.5 & & & 17.78 (0.11) & & ROTSE-IIIb \\
Dec 06, 2005 & 2453710.59& +16.1 & & & 17.78 (0.10) & & ROTSE-IIId \\
Dec 06, 2005 & 2453711.00& +16.5 & & & 17.85 (0.10) & & ROTSE-IIIb \\
Dec 07, 2005 & 2453711.99& +17.5 & & & 17.87 (0.09) & & ROTSE-IIIb \\
Dec 09, 2005 & 2453713.58& +19.1 & & & 17.73 (0.10) & & ROTSE-IIId \\
Dec 09, 2005 & 2453713.99& +19.5 & & & 17.76 (0.09) & & ROTSE-IIIb \\
Dec 10, 2005 & 2453714.58& +20.1 & & & 18.07 (0.12) & & ROTSE-IIId \\
Dec 10, 2005 & 2453714.70& +20.2 & 18.66 (0.10) & 18.28 (0.05) & 18.12 (0.05) & 17.90 (0.07) & AFOSC \\
Dec 10, 2005 & 2453714.99& +20.5 & & & 18.13 (0.16) & & ROTSE-IIIb \\
Dec 11, 2005 & 2453715.57& +21.1 & & & 18.13 (0.16) & & ROTSE-IIId \\
Dec 12, 2005 & 2453716.58& +22.1 & & & 18.20 (0.24) & & ROTSE-IIId \\
Dec 13, 2005 & 2453717.56& +23.1 & & & $>$18.06  & & ROTSE-IIId \\
Dec 17, 2005 & 2453721.95& +27.5 & & & $>$18.19  & & ROTSE-IIIb \\
Dec 19, 2005 & 2453723.95& +29.5 & & & $>$18.38  & & ROTSE-IIIb \\
Dec 20, 2005 & 2453724.96& +30.5 & & & $>$18.35  & & ROTSE-IIIb \\
Dec 23, 2005 & 2453727.62& +33.1 & & 19.90 (0.07) & & & AFOSC \\
Dec 23, 2005 & 2453727.64& +33.1 & 20.33 (0.08) & 19.90 (0.05) & 19.70 (0.10) & 19.38 (0.10) & AFOSC \\
Dec 24, 2005 & 2453728.78& +34.3 & 20.39 (0.20) & 19.94 (0.10) & 19.79 (0.20) & 19.47 (0.12) & RATCAM \\
Dec 27, 2005 & 2453731.95& +37.5 & & & $>$18.78  & & ROTSE-IIIb \\
Dec 31, 2005 & 2453735.94& +41.4 & & & $>$19.29  & & ROTSE-IIIb \\
Jan 03, 2006 & 2453738.60& +44.1 & $>$20.99 & 20.57 (0.18) & 20.21 (0.29) & 20.01 (0.31) & RATCAM \\
Jan 06, 2006 & 2453741.92& +47.4 & & & $>$19.61  & & ROTSE-IIIb \\
Jan 07, 2006 & 2453742.92& +48.4 & & & $>$19.57  & & ROTSE-IIIb \\
Jan 10, 2006 & 2453745.99& +51.5 & & & $>$19.32  & & ROTSE-IIIb \\
Jan 22, 2006 & 2453757.70& +63.2 & $>$22.39  & & & & RATCAM \\
Jan 24, 2006 & 2453759.87& +65.4 & & & $>$19.43  & & ROTSE-IIIb \\
Jan 30, 2006 & 2453765.69& +71.2 & $>$22.98 & $>$22.96 & 21.85 (0.19) & $>$21.85 & RATCAM \\
Mar 01, 2006 & 2453795.74& +101.2 & & & $>$19.38  & & ROTSE-IIIb \\
Mar 25, 2006 & 2453819.76& +125.3 & & & $>$19.48  & & ROTSE-IIIb \\
Apr 30, 2006 & 2453855.77& +161.3 & & & $>$19.66  & & ROTSE-IIIb \\
May 24, 2006 & 2453879.77& +185.3 & & & $>$19.50  & & ROTSE-IIIb \\
Jun 16, 2006 & 2453902.72& +208.2 & & & $>$19.03  & & ROTSE-IIIb \\
\hline
\end{tabular}
\end{table*}

Most of our data are unfiltered CCD images obtained by the 
Texas Supernova Search team \citep{qui06}, using a 0.45-m ROTSE-IIIb robotic telescope 
at the McDonald Observatory, which is one of the four telescopes of the ROTSE collaboration\footnote{http://rotse.net/}
 used for the optical monitoring of $\gamma$-Ray Burst (GRB) afterglows. 
Additional unfiltered points were obtained with the ROTSE-IIId telescope
of the Turkish National Observatory of Bakirlitepe (Turkey).
Both ROTSE-IIIb and ROTSE-IIId have Marconi CCDs covering a very large field (1.85$\times$1.85 square
degrees). Deep reference images (limiting magnitude $\sim$21) were obtained combining 
with a median filter several frames acquired during the period between 2004 Dec. 15 and 2005 Jan. 16.
The SN magnitudes were computed after subtracting the reference images and then
rescaled to the R-band magnitudes using a sequence of stars from the USNO-B1.0 catalogue. 
Since SN 2005la exploded quite close (7.5$\arcmin$) to a 5th magnitude star (LS Com), the SN photometry
was marginally affected by the presence of wings of light of this luminous source. More detail on the 
techniques used to derive the ROTSE-III magnitudes can be found in \citet{qui07}. 

Additional multicolour photometry was obtained using the 1.82-m Copernico Telescope of Mt. Ekar (Asiago, Italy)
equipped with AFOSC and the 2-m Liverpool Telescope (La Palma, Canary Islands) with RATCAM. The data reduction 
steps were performed with standard techniques (see e.g. \cite{pasto07c}), using IRAF tasks.
The SN magnitudes were obtained with the point-spread-function (PSF) fitting
technique after the subtraction of galaxy templates obtained on Feb. 2007.
The instrumental SN magnitudes were calibrated using standard fields by \citet{lan92}, and then
checked (and, if necessary, rescaled) with reference to a sequence of 5 stars in the SN vicinity 
(Fig. \ref{fig1} and Tab. \ref{tab1}).  
 The  photometric data of SN~2005la are reported in Tab. \ref{tab2}. 

\begin{figure*}
\centering
\includegraphics[angle=270,width=16cm]{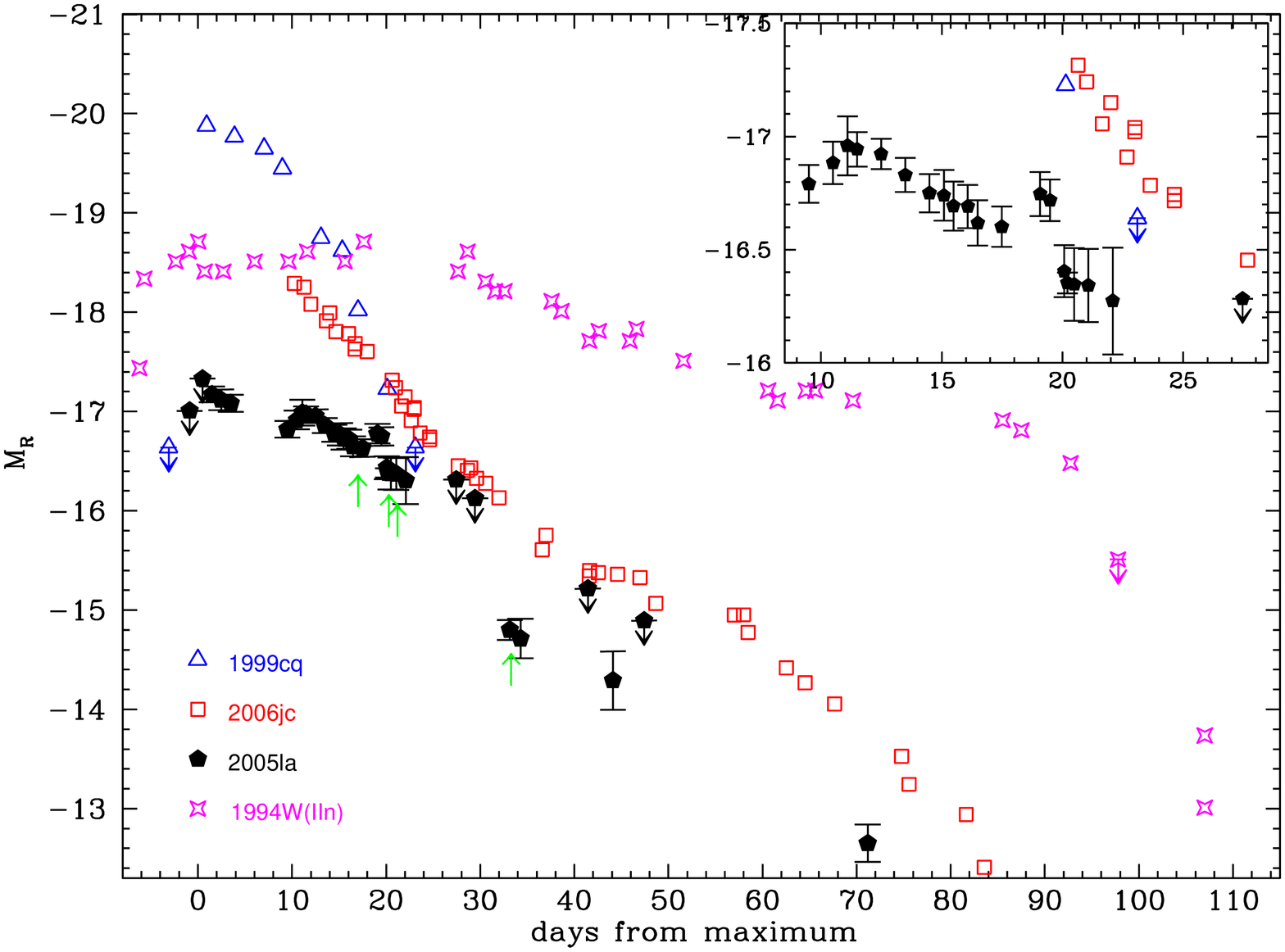}
\caption{R-band light curve of  SN 2005la compared with those of the type Ibn SNe
2006jc and 1999cq \protect\citep{pasto07,pasto07b,fol07,mat00}, and the type IIn SN 1994W (unfiltered, V-band and R-band photometry 
from \protect\cite{soll98}). Remarkably, SN~2005la shows a non-monotonic luminosity decay.
Unfiltered ROTSE-III magnitudes from the Texas Supernova Search have been
scaled to the R-band magnitudes. The most significant ROTSE-III limits are also reported.
A blow-up of the same light curves between 8-28 days is also shown  (top-right insert). 
Green arrows also mark the phases of the spectra of SN~2005la presented in this paper.
\label{fig2}}
\end{figure*}
  
Although the first ROTSE-IIId pre-discovery observation at Julian Date (JD) = 2453693.62 is not very deep, 
it provides a magnitude limit of R = 17.5, which is about 0.2 magnitudes fainter than the first, real detection. 
This is an indication that the SN was probably still rising in luminosity.
This limit, although not very stringent (see Tab. \ref{tab2}), is useful to constrain the epoch of the first 
maximum light, with an uncertainty of a couple of days. However, because of 
the non-monotonic behaviour of the light curve of SN~2005la at later phases, we cannot exclude the existence of
further, earlier luminosity peaks and, as a consequence, we cannot definitively constrain 
the explosion epoch. However, the relatively blue colour
of the early spectra (Sect. \ref{spec}) suggests that the SN was discovered rather young. 
Throughout this paper we will adopt JD = 2453694.5$\pm$1.5 as an indicative epoch of maximum light.

The R-band absolute light curves of SNe 2005la,  1999cq \citep{mat00} and 2006jc \citep{pasto07,fol07} 
are compared in Fig. \ref{fig2} with that of the type IIn SN 1994W \citep{soll98,chu04}. 
For computing the absolute light curves of SN 2006jc and SN 1999cq, we used the same parameters as reported 
in  Tab. 1 of \citet{pasto07b}.
The light curves of SN~2005la and 
SNe Ibn have a very rapid luminosity evolution compared with SN~1994W, which shows a sort of plateau lasting
about 3 months, followed by a very steep luminosity decline. 
Other SNe IIn, e.g. SN 1988Z \citep{are99} and SN 1995G \citep{pasto02}, show even more slowly evolving
light curves, without any evidence of a significant luminosity drop over timescales of a few years. 
   
\begin{table*}
\small
\footnotesize
\caption{Basic information of the spectra of SN 2005la.} \label{tab3}
\begin{tabular}{ccccccc} \\
\hline\hline
Date & JD & Phase (days) & Intrumental configuration & Integration time (s) & Range (\AA) & Resolution (\AA)\\
\hline
Dec. 07, 2005 & 2453711.5  & 17.0 & HET+LRS+G1 & 1200 & 4150-9200 & 17 \\
Dec. 10, 2005 & 2453714.8 & 20.3 & A1.82+AFOSC+gm4 & 1690 & 3600-7800 & 25 \\
Dec. 11, 2005 & 2453715.7 & 21.2 & A1.82+AFOSC+gm4 & 7200 & 3600-7800 & 25 \\
Dec. 23, 2005 & 2453727.8 & 33.3 & A1.82+AFOSC+gm4 & 6610 & 3550-7800 & 24 \\ 
\hline
\end{tabular}
\end{table*}

A large luminosity deficit at the optical wavelengths at late phases
can be attributed either to an extremely low amount of $^{56}$Ni
synthesized by the SN explosion or, alternatively, to an increase in
the extinction as a result of new dust being produced by the SN.  A
consequence is that increased emission at infrared wavelengths can be
expected, as a result of absorption and re-radiation of the
SN light by dust grains. Interestingly, SN 2006jc developed a
strong excess at near-infrared (NIR) wavelengths  \citep{ark06,smi07,sak07,dica07,seppo07},
with a simultaneous rapid decline of the optical light
curves. Condensation of dust in the ejecta of a few core-collapse SNe
has been observed at late times \citep[see][and references therein]{mei07}. 
Furthermore, for the type IIn SN 1998S it was
suggested that after 300 days new dust condensed in a cool, dense
shell (CDS) produced by the impact of the SN ejecta with a dense CSM \citep{poz04}.
A similar mechanism also occurred in SN
2006jc, but producing new dust at a much earlier epoch than in the
case of SN 1998S \citep[see discussion in][]{smi07,seppo07}. 
As in the case of SN 2006jc, newly-condensed CDS dust might
be responsible for the rapid decline of the optical light curves seen
also in other SNe Ibn (including SN 2005la). However, due to the lack of
NIR observations, it is impossible to verify if a strong IR
excess is a common characteristic of all these events.

\begin{figure}
\centering
\includegraphics[width=8.8cm]{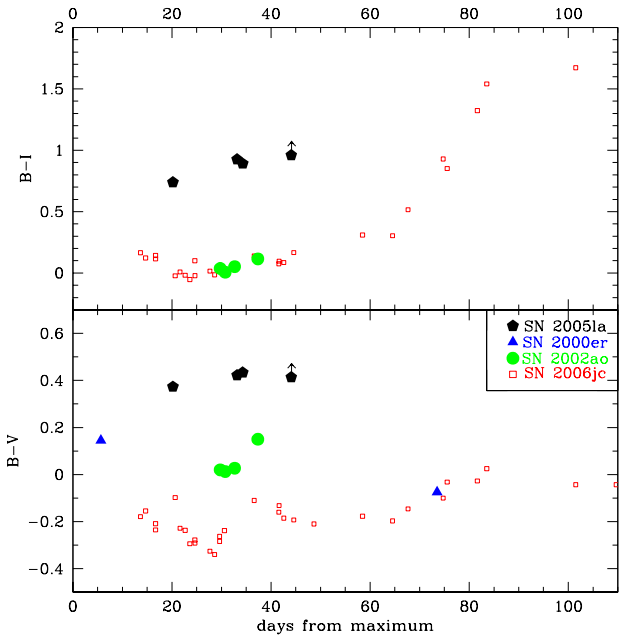}
\caption{B-I (top) and B-V (bottom) colour curves of SN~2005la,
compared with those of a few type Ibn SNe. We remark that the B-I colour is less
affected by the flux contamination of the narrow He I circumstellar lines than 
the B-V colour. Differences in the intensity of the He I circumstellar lines 
may explain why SNe Ibn show a very similar B-I colour evolution, 
but, simultaneously, rather different B-V colours. The amount of interstellar 
extinction adopted for the SNe Ibn sample is the same as in Tab. 1 of \protect\cite{pasto07b}.
\label{fig2b}}
\end{figure}
                                                                       
Although the rapid decline of the light curve of SN 2005la is similar to SN~2006jc, the overall luminosity evolution 
is substantially different. With the adopted scale in phase, its luminosity peak, M$_R$ = -17.2, is fainter 
(by about 1 magnitude) than that of SN~2006jc, and its R-band light curve has a non-monotonic
behaviour. In fact, after the initial post-discovery decline, the light curve shows at least one
secondary maximum at about 12 days post-discovery. Moreover, a further, luminosity peak is possibly visible 
at the phase of $\sim$ 3 weeks, although we cannot definitely exclude this as being an artifact due to measurement errors.
However, it is remarkable that  before and after this apparent peak, the spectra evolve significantly, 
showing broader spectral lines and less evident P-Cygni absorptions (see Sect. \ref{spec}). This may
indicate a new, stronger episode of ejecta-CSM interaction.
After that, for about two months (between phase of 20 and 80 days) the R-band magnitude decline 
is very pronounced ($\Delta$M$_R \sim$ 4.2 magnitudes).

In Fig. \ref{fig2b} the B-I and B-V colour curves of SN~2005la are shown, together with those of the type Ibn SNe 2000er, 
2002ao and 2006jc. Unfortunately, except for SN~2006jc, the colour information for SNe Ibn is incomplete.
However, from available data, the B-I colour evolves in a similar fashion in all SNe Ibn, with very little 
colour evolution during the first two months (B-I always ranges between 0 and 0.3 magnitudes). However, while in 
SN~2006jc B-V remains between -0.3 and -0.1 during the first couple of months, SN~2002ao has a significantly 
redder B-V colour, probably due to the presence of more prominent emission features in the V- and R-band regions 
\citep[see Fig. 4 of][]{pasto07b}. 
On the other hand, SN~2000er roughly follows the B-V colour evolution of SN~2006jc. 
Unfortunately the colour evolution of SN~2005la is not well sampled. 
Nevertheless, 3-4 weeks after maximum, this SN appears to be significantly redder 
(B-I $\approx$ 0.9; B-V $\approx$ 0.4) than any type Ibn SN.

While any colour information is missing for SN~2005la at about 2 months after maximum, 
SNe Ibn become much redder (see Fig. \ref{fig2b} top, reaching B-I $\approx$ 1.6 one month later). 
This is because starting from $\sim$2 months past maximum the contribution in flux at the blue 
wavelengths decreases significantly in type Ibn SNe,
while a cooler black body component due to dust is responsible for the flux excess at the red
wavelengths \citep[at least in SN~2006jc, see][]{smi07,seppo07}.   

\section{Spectra} \label{spec}

\begin{figure}
\centering
\includegraphics[width=8.9cm]{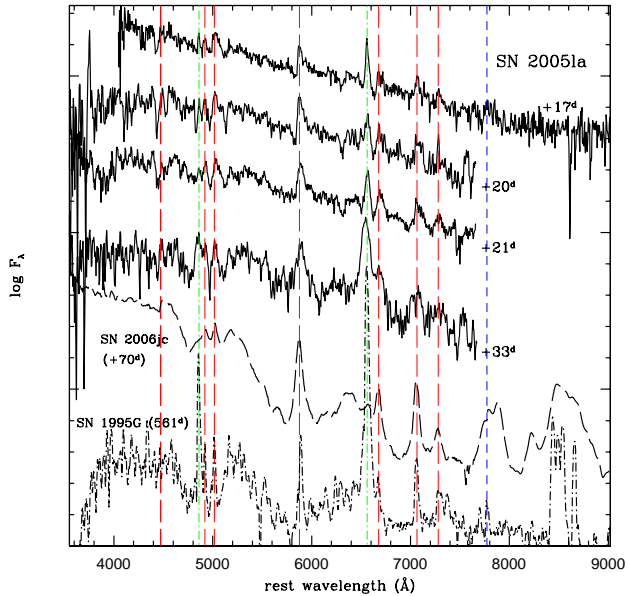}
\caption{Spectroscopic evolution of SN~2005la. A spectrum of the 
type Ibn SN~2006jc \protect\citep{pasto07} at a phase of 70 days and a very late spectrum of the type
IIn SN 1995G \protect\citep[phase 561 days,][]{pasto02} 
are also shown as a comparison. Vertical red
dashed lines mark the position of the main He I features, while the dot-dashed
green lines mark the rest wavelength position of H$\alpha$ and H$\beta$. 
The blue dashed line marks the position of the OI 7774\AA~ feature. 
All spectra are at the host galaxy rest wavelength. \label{fig3}}
\end{figure}

\begin{table*}
\centering
\footnotesize
\caption{FWHM velocities for the strongest H and He I lines in the spectra of SN~2005la. The expansion velocities
derived from the position of the P-Cygni minima of a few lines are also reported, labelled with letters, below 
the Table.} \label{tab4}
\begin{tabular}{cccccccc} \\
\hline
Date &  Phase & H$\alpha$ & H$\beta$ & He I $\lambda$4922 & He I $\lambda$5876 & He I $\lambda$6678 &He I $\lambda$7065 \\
 & (days) & (km s$^{-1}$) &  (km s$^{-1}$) &  (km s$^{-1}$)&  (km s$^{-1}$) &  (km s$^{-1}$) &  (km s$^{-1}$)  \\ 
\hline
Dec. 07, 2005 & 17.0 & 1550$\pm$170 & 1480$\pm$110 & 2140$\pm$240 & 2160$\pm$250$^{(a)}$ & 2020$\pm$480$^{(b)}$ & 2160$\pm$200 \\
Dec. 10, 2005 & 20.3 & 1970$\pm$490  &  &  &  3160$\pm$400 &  3100$\pm$990 & 3120$\pm$740\\
Dec. 11, 2005 & 21.2 & 1950$\pm$360  & 1870$\pm$270 & & 3160$\pm$400 & 3140$\pm$380$^{(c)}$  & 3150$\pm$480\\
Dec. 23, 2005 & 33.3 & 4200$\pm$260  & 4170$\pm$640 & & 4190$\pm$460 & 4140$\pm$670 & 4150$\pm$990\\ 
\hline
\end{tabular}

$^{(a)}$  2170$\pm$210; $^{(b)}$  1960$\pm$310;  $^{(c)}$  3210$\pm$250. 
\end{table*}

The very rapid evolution of SN~2005la and its moderate apparent luminosity prevented us
from performing extensive spectroscopic monitoring, and only 4 spectra were collected. 
The first spectrum was obtained with 
the 9.2-m Hobby-Eberly Telescope (HET) of the McDonald Observatory (University of Texas, US), 
equipped with the Marcario Low-Resolution Spectrograph (LRS) on Dec. 7. 
Additionally, 3 spectra were obtained later using
the 1.82-m Copernico Telescope with AFOSC. 
More information on the 4 spectra is reported in Tab. \ref{tab3}, while
the complete spectral sequence is shown in Fig. \ref{fig3}.

\begin{figure}
\centering
\includegraphics[width=8.6cm]{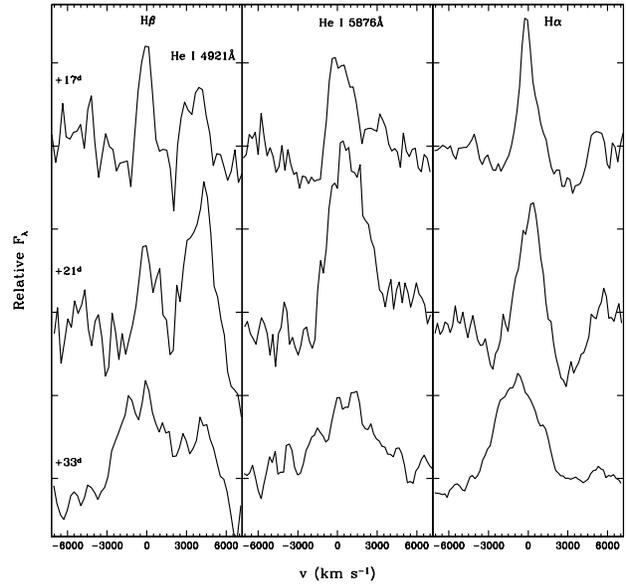}
\caption{Evolution of the profile of individual H and He I lines. \label{fig3b}}
\end{figure}

In Fig. \ref{fig3} the spectra of SN~2005la are also compared with a spectrum of the type Ibn SN~2006jc 
at phase +70$^d$ \citep{pasto07} and a very late spectrum of the type IIn SN~1995G \citep{pasto02}. This comparison
highlights a rough similarity between the spectra 
of SN~2005la (especially at 20-21$^d$ after the discovery) and that of  SN~2006jc. 
However, H$\alpha$ is prominent in SN~2005la (although much weaker than in SN 1995G). 
Its H$\alpha$ is comparable with the most prominent He I lines (i.e. $\lambda$5876 and $\lambda$7065) 
at early epochs and is stronger than the He I lines only in the last spectrum. Fig. \ref{fig3b} shows the evolution 
of the profiles of H$\beta$ (+ He I $\lambda$4921), He I $\lambda$5876 and H$\alpha$ 
at three selected epochs.
The relative strengths of the H and He I lines may change as a result of abundance and, possibly, 
ionization effects. The latter is mostly expected in objects whose ejecta are interacting with a CSM. 

Unlike SN 2005la, the spectrum of SN~2006jc shows a very weak (practically invisible) H$\alpha$ during the first 3 months
\citep{pasto07,fol07}. 
This line becomes more prominent only at very late epochs \citep{smi07,pasto07b,seppo07}.
Instead, in type IIn SNe, a narrow (typically with v$_{FWHM} \leq$  1000 km s$^{-1}$) H$\alpha$ dominates in strength at any phase
over the He I lines.

In Tab. \ref{tab4} we report the velocities of SN~2005la as derived from measuring the full-width-half-maximum (FWHM) 
of a number of lines (H and He I). While the He I lines show a clear P-Cygni profile, this is not obvious for 
the H lines. The velocities derived from the position of the P-Cygni minima of the He I lines 
are quite consistent ($\sim$ 2000-2200 km s$^{-1}$, see also Tab. \ref{tab4}) with the FWHM velocities (v$_{FWHM}$) of the same lines. 
We find that in the 7th December spectrum the v$_{FWHM}$ derived for the H lines 
is about 1500 km s$^{-1}$ , while it is slightly higher for the He I lines (v$_{FWHM}$ $\approx$ 2150 km s$^{-1}$).
These lines are probably produced in an
unshocked CSM surrounding the SN. Once the SN ejecta reach the CSM, 
this material is expected to accelerate. A bump near 7800\AA~is tentatively identified as a proper SN
ejecta feature, probably O I $\lambda$7774. The low S/N does not allow us to properly measure
the velocity of this feature, but (despite the uncertainty) it is significantly higher (v$_{FWHM}$ = 4900$\pm$1200)
than the other lines identified in the December 7th spectrum. This velocity can be considered as representative
for the ejecta velocity.

The two subsequent spectra (December 10 and 11) show that the line velocities are increasing 
(v$_{FWHM}$ $\approx$ 1900 km s$^{-1}$ for the H lines; v$_{FWHM}$ $\approx$ 3150 km s$^{-1}$ for the He lines), indicating
an evident acceleration of the He- and, more marginally, of the H-rich CS region.
In the last spectrum of SN~2005la, obtained about two weeks later, all lines are in pure emission (see Fig. \ref{fig3b}), 
as expected when driven by the high-energy radiation resulting from the 
interaction of the ejecta with the dense CSM close to the SN.
Moreover, the velocity of both the H and He I lines is much higher, being
v$_{FWHM}$ $\approx$  4200 km s$^{-1}$, which is close to that estimated for the O I $\lambda$7774 line.
No extended wings are observed for the emission features, as one would expect if the line
broadening was a result of scattering of photons of the narrow lines by the thermal
electrons of the circumstellar gas \citep[as was observed in SN 1998S,][]{chu01}.

It is remarkable that the absorption features in the region between 4600 and 5200\AA~(possibly due to Fe II lines and
detected with comparable widths also in SNe Ib/c), the overall blue colour and the drop of the flux beyond 5400\AA, 
seem to be common characteristics of most interacting core-collapse SNe (Fig. \ref{fig3}). 

\section{Searching for major pre-SN outbursts} \label{outburst}

SN~2005la shows clear evidence of the presence of CSM produced by the precursor star before
the SN explosion. The SN spectra indicate that this material is rich in both H and He.
However there are some spectroscopic indications that the inner CSM is more He-rich, while H is more abundant
in more external regions of the CSM. From the available data, it is very difficult to understand if this material
was produced through a dense, continuum wind or via major outburst episodes similar to those observed 
in a number of  LBVs \citep[or, alternatively, in luminous outbursts of WR stars analogous to the transient 
UGC 4904-V1 which preceded by two years the explosion of SN~2006jc, see][]{pasto07}.

\begin{figure}
\includegraphics[width=8.65cm]{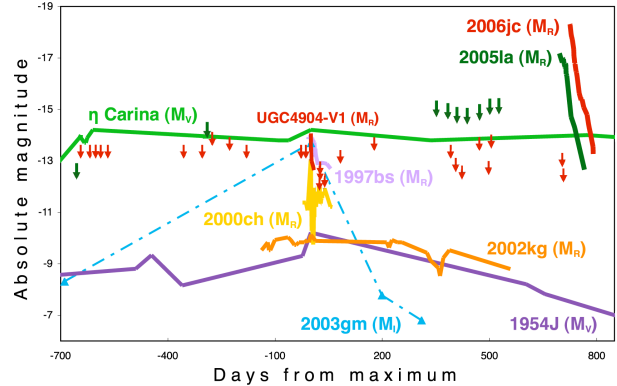}
\caption{Pre-SN detection limits and R-band absolute light curve of SN~2005la compared
with several well-known outbursts of LBVs and the LBV-like transient UGC 4904-V1 which occurred
two years before the explosion of SN~2006jc (its light curve is also shown in Figure). The phases
are relative to the LBV outburst maximum light.
The figure is similar to that shown in \protect\citet[][see references therein]{pasto07} with,
in addition, pre- and post-explosion data of SN~2005la. Lacking the detection of a giant outburst 
preceding SN 2005la, the pre-SN detection limits and the light curve of SN 2005la are shown
in an arbitrary temporal scale. \label{fig4}}
\end{figure}

Up to now, it was commonly accepted that core-collapse explosions occur substantially ($\sim$ a few times 10$^{5}$ years) 
after the last major mass-loss episode. However, the precedent of SN~2006jc should give impulse to the efforts of finding evidence 
for similar luminous outbursts occurring only a few years before the explosion of interacting core-collapse SNe. Unfortunately, 
the host galaxy of SN~2005la was not frequently monitored in the past, and very few images are currently available in public 
archives. We analysed B and Sloan r band images of KUG 1249+278 (with exposure times of 3$\times$300s and 600s, respectively)
obtained on 2002 March 10th with the 2.5-m Isaac Newton Telescope in La Palma (Canary Islands, Spain). 
The data have been downloaded from the ING archive\footnote{http://casu.ast.cam.ac.uk/casuadc/archives/ingarch}.
Another set of images were obtained making use of the SMOKA\footnote{http://smoka.nao.ac.jp/servlet/search, 
\protect\citet{baba02} } science
archive. The host of SN~2005la was observed at the 1.05-m Schmidt Telescope of the Kiso Observatory (University of Tokyo, Japan)
on 2003 March 11th (4$\times$300s and 3$\times$300s in the V and R bands, respectively).
Finally, the Texas SN Search has intensively (almost daily) monitored the region of KUG 1249+278 (from November 20, 2004 to 
July 7, 2005), but no evidence of pre-SN outbursts was found (Fig. \ref{fig4}).
This is not really unexpected since the deepest magnitude limits registered for the ROTSE-IIIb telescopes
are near magnitude 20. With the assumptions on the distance and extinction discussed in
Sect. \ref{general}, only outbursts more luminous than M$_R$ $\sim$ -14.5 would be detected,
which is an absolute magnitude very close to the brightest eruptions reported in Fig. \ref{fig4}, and only
$\sim$2.7 magnitudes fainter than the peak magnitude registered for SN~2005la.
Therefore, the analysis of the pre-explosion images do not provide any useful information
on the mass loss processes that occurred in the last years of life of the progenitor of SN~2005la.

\section{What was SN~2005la?} \label{discussion}

The spectroscopic properties of SN~2005la, especially the peculiar intensity ratio between H$\alpha$ and
the most prominent He I lines ($\lambda$5876 and $\lambda$7065) suggest that this object 
could be classified as lying somewhere between the 
classical type IIn SNe \citep{sch90} and the type Ibn events \citep{pasto07,pasto07b}. 
However, the rapidly evolving light curve of SN~2005la, and the very fast decline of the optical
luminosity (see Fig. \ref{fig2}) is more typical for the family of SN~2006jc. 

There now appears to be strong observational evidence that normal type II-P SNe have 
red supergiant progenitors which have fairly moderate initial masses
\citep{sma04,2007ApJ...661.1013L}. It is interesting to note that there are no discoveries of 
any red supergiants with masses definitely more than 20M$_{\odot}$. 
The case of the blue supergiant exploding as SN 1987A is 
still not well understood from the stellar evolutionary point of view, but
binarity is the most likely explanation \citep{1992PASP..104..717P}. Hence one might conjecture
that higher mass stars explode as some other type of event. 

There is a reasonable probability that Wolf-Rayet (WR) stars of type
WN and WC/O are the direct progenitors of at least some SNe~Ib and Ic. Recent modelling of
the light curves of type Ic SNe have yielded ejecta masses (C and O dominated)
that have some overlap with  the core masses of WC stars determined from 
luminosity measurements and stellar evolutionary models. For example, 
ejecta masses of 2.5-5M$_{\odot}$ for SN 2002ap \citep{2002ApJ...572L..61M} 
and 3.5-8M$_{\odot}$ for SN 2004aw \citep{tau06} were determined, 
while \cite{2002A&A...392..653C} suggest masses of 7-19M$_{\odot}$ for WC stars
in the Large Magellanic Cloud (LMC) and the Galaxy. While these mass ranges
just about overlap, the fact that the explosion of SN 2002ap was asymmetric and lack of
a large sample of SN Ib/c ejecta or progenitor mass estimates still leaves open
the question of the origins of these events.

There is also mounting evidence that 
type Ib/c SNe are associated with environments which show stronger H$\alpha$ 
emission and have higher surface brightness than do the environments of type II-P SNe \citep{2006A&A...453...57J,kelly08}. 
This would suggest the progenitors are more likely from higher mass 
stars, and hence above 15-20M$_{\odot}$. Although
rather unexpectedly the spatial type Ib SN distribution of \citet{kelly08} resembles the type II SN distribution.
There is one possible detection of a 
type IIn SN progenitor, that of SN 2005gl \citep{gal07}, although the object is 
exceptionally bright for a single star and it remains uncertain if it is
actually one of the most luminous stars that can exist, a very high-mass
LBV or if it is actually an unresolved stellar cluster. 


In the case of the type IIn SN 2005gl, the ejecta started to interact early 
with the material likely lost through major eruptions a short time before the explosion of the star.
From this point of view, there is a remarkable analogy with the sequence of events that led to
SN 2006jc in UGC 4904, as a luminous pre-SN outburst followed two years later by
the real SN explosion \citep{ita06,pasto07}. 
In such a case, if the precursor of SN 2006jc was a single massive star, it was responsible 
for the outburst observed on October 2004 \citep{pasto07}, in which a 
significant amount of material was likely expelled. SN 2006jc was monitored in detail 
for several months after October 2006, and the spectra showed
relatively narrow He I emission lines, indicating that the SN ejecta were interacting  
with a slower moving He-rich CSM. 
As a consequence, the 2004 luminous outburst was probably produced by a WR star 
transiting between the WN and WC/O 
phases \citep[e.g.][]{fol07,tom07}. 
Perhaps {\em all} precursor stars of the interacting type Ibn SNe 
are rare objects \citep{pasto07b} which lie in the region between more typical WN and WC/O stars 
(which should produce SNe Ib and SNe Ic, respectively). One might draw a parallel between 
this argument and that used to relate the LBV progenitors of SNe IIn 
to the RSG/BSG progenitors of non-interacting SNe II. 

In this context, SN~2005la is unique, because it shares properties with both SNe IIn and SNe Ibn. 
Hence we suggest that 
SN~2005la might have been produced by the explosion of a star which was transiting from the LBV to 
the WR (WN-type) phase. From the analysis of the SN properties, the young WN precursor of SN~2005la
should have suffered a major mass loss episode (or possibly several episodes) 
a few years before the SN explosion.  One would expect that 
during these episodes a significant amount of He-rich material 
(with residual H) was lost. The detection of outbursts similar to that seen for SN 2006jc and 
other similar SNe would be an exciting discovery as it would hint that the instability 
may be directly related to the imminent core-collapse. Unfortunately,
the analysis of the available pre-SN archive images for SN 2005la (see Sect. \ref{outburst})
does not provide any direct confirmation for this scenario although, of course, it does
not exclude it. 
Wide field transient searches which sample low-mass and hence low metallicity galaxies 
are the obvious way forward to discover more of these peculiar and interesting SNe. For
example the blank-field strategy of the Texas Supernova Search has discovered a large number of 
peculiar SNe \citep[e.g. the weird SN~2005ap,][]{qui07b}. 
The future Pan-STARRS project \citep{young08} will have the capability of increasing these
discoveries further and more extensive monitoring of the events will help to shed light 
on the post-main-sequence evolution of the most massive stars.
\\

\section*{Acknowledgments}
SM acknowledges financial support from the Academy of
Finland (project: 8120503). \\
The paper is based on observations collected at the 0.45m ROTSE-IIIb robotic
telescope and the Hobby-Eberly Telescope at the McDonald Observatory (University of Texas, US), the 0.45m ROTSE-IIId robotic
telescope of the Turkish National Observatory (Bakirlitepe, Turkey), the 
Copernico 1.82-m Telescope of the INAF-Asiago Observatory (Mt. Ekar, Italy)
and the Liverpool Telescope (La Palma, Spain).
We thank the ROTSE collaboration, the Hobby-Eberly Telescope staff, particularly B. Roman, V. Riley, and S. Rostopchin,
and the support astronomers of the Liverpool Telescope for performing the 
observations of SN~2005la.  \\
This paper makes use of data obtained from the Isaac Newton Group Archive which is 
maintained as part of the CASU Astronomical Data Centre at the Institute of Astronomy, Cambridge.
This work is also based in part on data collected at the Kiso observatory (University of Tokyo) and 
obtained from the SMOKA science archive, which is operated by the Astronomy Data Center, 
National Astronomical Observatory of Japan. This research has made use of the NASA/IPAC Extragalactic
Database (NED) which is operated by the Jet Propulsion Laboratory,
California Institute of Technology, under contract with the National
Aeronautics and Space Administration. 
We acknowledge the usage of the HyperLeda database (http://leda.univ-lyon1.fr).\\


\begin{thebibliography}{99}
\bibitem[Antilogus et al. (2007)]{NSF07} Antilogus, P.  et al. 2007, CBET 941
\bibitem[Anupama et al. (2001)]{anu01} Anupama, G. C., Sivarani, T., Pandey, G. 2001, A\&A, 367, 506
\bibitem[Aretxaga et al. (1999)]{are99} Aretxaga, I., Benetti, S., Terlevich, R. J., Fabian, A. C., 
Cappellaro, E., Turatto, M., Della Valle, M. 1999, MNRAS, 309, 343
\bibitem[Arkarov et al. (2006)]{ark06} Arkharov, A., Efimova, N., Leoni, R., Di Paola, A., Di Carlo, E., Dolci, M. 2006,
Astron. Tel. 961 
\bibitem[Baba et al. (2002)]{baba02} Baba, H. et al. 2002, ADASS XI, eds. D. A. Bohlender, D. Durand $\&$ T. H. Handley, 
ASP Conference Series, Vol. 281, 298
\bibitem[Benetti et al. (1999)]{ben99} Benetti, S., Turatto, M., Cappellaro, E., Danziger, I. J., Mazzali, P. A.
1999, MNRAS, 305, 811
\bibitem[Benetti et al. (2007)]{ben06} Benetti, S., Cappellaro, E., Turatto, M., Taubenberger, S., Harutyunyan, A., 
Valenti, S. 2006, ApJ, 653L, 129
\bibitem[Chugai et al. (1995)]{chu95} Chugai, N. N., Danziger, I. J., Della Valle, M.
1995, MNRAS, 276, 530
\bibitem[Chugai et al. (2001)]{chu01} Chugai, N. N. 2001, MNRAS, 326, 1448
\bibitem[Chugai et al. (2004)]{chu04} Chugai, N. N. et al. 2004, MNRAS, 352, 1213
\bibitem[Crowther et al.(2002)]{2002A&A...392..653C} Crowther, P.~A., 
Dessart, L., Hillier, D.~J., Abbott, J.~B., Fullerton, A.~W.\ 2002, A\&A, 392, 653 
\bibitem[Di Carlo et al. (2008)]{dica07} Di Carlo, E. et al. 2008, ApJ in press (arXiv:0712.3855)
\bibitem[Dopita et al. (1984)]{dop84} Dopita, M. A., Cohen, M., Schwartz, R. D., Evans, R. 1984, ApJ, 287L, 69
\bibitem[Ercolano et al. (2007)]{erc07} Ercolano, B., Barlow, M. J., Sugerman, B. E. K. 2007, MNRAS, 375, 753
\bibitem[Fassia et al. (2001)]{fas01} Fassia, A. et al. 2001 MNRAS, 325, 907
\bibitem[Filippenko et al. (2005)]{fil05} Filippenko, A. V., Foley, R. J., 
Matheson, T. 2005, IAU Circ. 8639, 2
\bibitem[Foley et al. (2007)]{fol07} Foley, R. J., Smith, N., Ganeshalingam, M., 
Li, W., Chornock, R., Filippenko, A. V. 2007, ApJ, 657L, 105
\bibitem[Fransson et al. (2002)]{fran02} Fransson, C. et al. 2002, ApJ, 572, 350 
\bibitem[Gal-Yam et al. (2007)]{gal07} Gal-Yam, A. et al. 2007, ApJ, 656, 372
\bibitem[Gerardy et al. (2000)]{ger00} Gerardy, C. L., Fesen, R. A., H\"{o}flich, P., Wheeler, J. C.
2000, AJ, 119, 2968
\bibitem[Gerardy et al. (2002)]{ger02} Gerardy, C. L. et al. 2002, ApJ, 575, 1007
\bibitem[Hamuy et al. (2003)]{ham03} Hamuy, M. et al. 2003, Nature, 424, 651
\bibitem[Hagen et al. (1997)]{heg97}   Hagen, H.-J., Engels, D., Reimers, D.
1997, A\&A, 324L, 29
\bibitem[Immler et al. (2007)]{imm07} Immler, S. et al. 2007, ApJ, 674L, 85
\bibitem[Itagaki et al. (2006)]{ita06} Itagaki, K. et al. 2006, IAU Circ. 8762
\bibitem[James \& Anderson(2006)]{2006A&A...453...57J} James, P.~A., 
Anderson, J.~P.\ 2006, A\&A, 453, 57 
\bibitem[Kelly et al. (2008)]{kelly08} Kelly, P. L., Kirshner, R. P., Pahre, M. 2007, ApJ submitted
(arXiv:0712.0430)
\bibitem[Kotak \& Vink (2006)]{kot06} Kotak, R., Vink, J. S. 2006, A\&A, 460L, 5 
\bibitem[Landolt (1992)]{lan92} Landolt, A. U. 1992 AJ, 104,340 
\bibitem[Leonard et al. (2000)]{leo00} Leonard, D. C., Filippenko, A. V., Barth, A. J., Matheson, T. 2000 ApJ, 536,
239L
\bibitem[Li et al.(2007)]{2007ApJ...661.1013L} Li, W., Wang, X., Van Dyk, 
S.~D., Cuillandre, J.-C., Foley, R.~J., Filippenko, A.~V.\ 2007, ApJ, 
661, 1013 
\bibitem[Liu et al. (2000)]{liu00} Liu, Q.-Z., Hu, J.-Y., Hang, H.-R., Qiu, Y.-L., Zhu, Z.-X., Qiao, Q.-Y. 2000, A\&AS,
114, 219L
\bibitem[Matheson et al. (2000)]{mat00} Matheson, T. et al. 2000, AJ, 119, 2303
\bibitem[Mattila et al. (2008)]{seppo07} Mattila, S. et al. 2008, MNRAS accepted (arXiv:0803.2145)
\bibitem[Mazzali et al.(2002)]{2002ApJ...572L..61M} Mazzali, P.~A. et al., 2002, ApJ, 572L, 61 
\bibitem[Meikle et al. (2007)]{mei07} Meikle, W. P. S, et al. 2007, ApJ, 665, 608
\bibitem[Ofek et al. (2007)]{ofek07} Ofek, E. O. et al. ApJ, 659L, 130
\bibitem[Pastorello et al. (2002)]{pasto02} Pastorello et al. 2002, MNRAS, 333,27
\bibitem[Pastorello et al. (2005)]{pasto05} Pastorello, A., Aretxaga, I., Zampieri, L., Mucciarelli, P., Benetti, S.
2005, {\sl 1604-2004: Supernovae as Cosmological Lighthouses}, ASP Conference Series, Vol. 342, Proceedings of the conference 
held 15-19 June, 2004 in Padua, Italy. Edited by M. Turatto, S. Benetti, L. Zampieri, and W. Shea. San Francisco: Astronomical 
Society of the Pacific, 2005, p.285
\bibitem[Pastorello et al. (2007a)]{pasto07} Pastorello, A. et al. 2007a, Nature, 447, 829
\bibitem[Pastorello et al. (2007b)]{pasto07c} Pastorello, A. et al. 2007b, MNRAS, 376, 1301
\bibitem[Pastorello et al. (2008)]{pasto07b} Pastorello, A. et al. 2008, MNRAS, accepted (arXiv:0801.2277)
\bibitem[Prieto et al. (2007)]{pri07a} Prieto, J. L. et al. 2007, AJ submitted (arXiv:0706.4088)
\bibitem[Prieto et al. (2008)]{pri07b} Prieto, J. L., Stanek, K. Z., Beacom, J. F. 2008, ApJ, 673, 999
\bibitem[Puckett et al. (2005)]{puc05} Puckett, T., Sostero, G., Quimby, R.,
Mondol, P. 2005, IAU Circ. 8639
\bibitem[Podsiadlowski(1992)]{1992PASP..104..717P} Podsiadlowski, P.\ 1992, PASP, 104, 717 
\bibitem[Pozzo et al. (2004)]{poz04}  Pozzo, M., Meikle, W. P. S., Fassia, A., Geballe, T., Lundqvist, P., Chugai, N. N., 
Sollerman, J. 2004, MNRAS, 352, 457
\bibitem[Quimby (2006)]{qui06} Quimby, R. M. 2006, Ph.D. thesis, Univ. Texas
\bibitem[Quimby et al. (2007a)]{qui07} Quimby, R. M., Wheeler, J. C., H\"oflich, P., Akerlov, C. W., Brown, P.
J, Rykoff, E. S. 2007a, ApJ, 666, 1093
\bibitem[Quimby et al. (2007b)]{qui07b} Quimby, R. M., Aldering, G., Wheeler, J. C., H\"oflich, P., Akerlov, C. W., Rykoff, E. S.
2007b, ApJ, 668L, 99
\bibitem[Sakon et al. (2007)]{sak07} Sakon, I. et al. 2007, ApJ submitted (arXiv:0711.4801)
\bibitem[Salamanca et al. (2002)]{sal02} Salamanca, I., Terlevich, R. J., Tenorio-Tagle, G. 2002, MNRAS, 330, 844
\bibitem[Sollerman et al. (1998)]{soll98} Sollerman, J., Cumming, R. J., Lundqvist, P. 1998, ApJ, 493, 933
\bibitem[Schlegel et al. (1998)]{sch98} Schlegel, D. J., Finkbeiner, D. P., Davis, M. 1998, ApJ, 500, 525 
\bibitem[Schlegel  (1990)]{sch90} Schlegel, E. M. 1990, MNRAS, 244, 269
\bibitem[Schlegel et al., (1996)]{sch96} Schlegel, E. M., Kirshner, R. P., Huchra, J. P., Schild, R. E. 1996,
AJ, 111, 2038
\bibitem[Smartt et al. (2004)]{sma04} Smartt, S. J. et al. 2004, Science, 303, 499 
\bibitem[Smith et al. (2007)]{smi06} Smith, N. et al. 2007, ApJ, 666, 1116
\bibitem[Smith et al. (2008)]{smi07} Smith, N., Foley, R.J., Filippenko, A. V. 2008, ApJ, 680, 568 
\bibitem[Stathakis \& Sadler (1991)]{sta91} Stathakis, R. A., Sadler, E. M. 1991, MNRAS, 250, 786
\bibitem[Sugerman et al. (2006)]{sug06} Sugerman, B. E. K. et al. 2006, Science, 313, 196
\bibitem[Taubenberger et al. (2006)]{tau06} Taubenberger, S. et al. 2006, MNRAS, 371, 1459
\bibitem[Tominaga et al. (2007)]{tom07} Tominaga, N. et al. 2007, ApJ submitted (arXiv:0711.4782)
\bibitem[Tremonti et al. (2004)]{tre04} Tremonti, C. A. et al. 2004,  ApJ, 613, 898
\bibitem[Trundle et al. (2008)]{tru08} Trundle, C., Kotak, R., Vink, J. S., Meikle, W. P. S. 2008, A\&A, 483L, 47
\bibitem[Turatto et al. (1993)]{tur93} Turatto, M., Cappellaro, E., Danziger, I. J., Benetti, S., Gouiffes, C., 
Della Valle, M. 1993 MNRAS, 262, 128
\bibitem[Young et al. (2008)]{young08} Young D. R. et al. 2008, A\&A submitted
\bibitem[Zampieri et al. (2005)]{zam05} Zampieri, L., Mucciarelli, P., Pastorello, A., Turatto, M., 
Cappellaro, E., Benetti, S. 2005, MNRAS, 364, 1419
\bibitem[Woosley et al. (2007)]{woo07} Woosley, S. E., Blinnikov, S., Heger, A. 2007, Nature, 450, 390
\end{thebibliography}
\end{document}